# Feasibility study of clinical target volume definition for soft-tissue sarcoma using muscle fiber orientations derived from diffusion tensor imaging

Authors: Nadya Shusharina[1,7], Xiaofeng Liu[2,7], Jaume Coll-Font[3,4,7], Anna Foster[3,4], Georges El Fakhri[2,7], Jonghye Woo[2,7], Thomas Bortfeld[1,7], Christopher Nguyen[3,4,5,6,7]

[1]Division of Radiation Biophysics, Department of Radiation Oncology, Massachusetts General Hospital, Boston, MA 02114, USA.

[2]Gordon Center for Medical Imaging, Department of Radiology, Massachusetts General Hospital, Boston MA 02114, USA.

[3]Cardiovascular Research Center, Massachusetts General Hospital, Charlestown, MA 02129, USA.

[4]A.A. Martinos Center for Biomedical Imaging, Massachusetts General Hospital, Charlestown, MA 02129, USA.

[5]Division of Health Science Technology, Massachusetts Institute of Technology, Cambridge, MA, USA

[6]Cardiovascular Innovation Research Center, Heart, Vascular, and Thoracic Institute, Cleveland Clinic, Cleveland, OH, USA

[7]Harvard Medical School, Boston, MA 02114, USA.

Corresponding author: Nadya Shusharina, nshusharina@mgh.harvard.edu
Address: 125 Nashua St., Suite 320, Boston, MA 02114, USA

**Abstract**

*Objective:* Soft-tissue sarcoma spreads preferentially along muscle fibers. We explore the utility of deriving muscle fiber orientations from diffusion tensor MRI (DT-MRI) for defining the boundary of the clinical target volume in muscle tissue.

*Approach:* We recruited eight healthy volunteers to acquire MR images of the left and right thigh. The imaging session consisted of (a) two MRI spin-echo-based scans, T1- and T2-weighted; (b) a diffusion weighted (DW) spin-echo-based scan using an echo planar acquisition with fat suppression. The thigh muscles were auto-segmented using CNN. DT-MRI data was used as a geometry encoding input to solve the anisotropic Eikonal equation with Hamiltonian Fast-Marching method. The isosurfaces of the solution modeled the CTV boundary.

*Main results:* The auto-segmented muscles of the thigh agreed with manually delineated with the Dice score ranging from 0.8 to 0.94 for different muscles. Anisotropy of the isosurfaces was compared across muscles with different anatomical orientations within a thigh, between muscles in left and right thighs of each subject, and between different subjects. Analysis showed a high degree of consistency across all comparisons. The distance from the GTV to the isosurface and the eigenvalues ratio are two controlling parameters for the extent and shape of the CTV.

*Significance:* Our feasibility study with healthy volunteers shows the promise of using muscle fiber orientations derived from diffusion weighted MRI data for automated generation of anisotropic CTV boundary in soft tissue sarcoma. Our contribution is significant as it is expected to lead to the improvements in the treatment outcomes of soft-tissue sarcoma patients undergoing radiotherapy and decrease amputation rate for a subset of patients. We expect such improvements to have a strong positive impact for the cancer centers with small volume of sarcoma patients.

**Introduction**

Combined surgery and radiotherapy is the primary treatment for soft-tissue sarcoma (STS) of the extremities (1-3) and shows improved local control and overall survival as compared to surgery alone, however, local tumor recurrence remains common (4-6). Improvement of radiotherapy treatment planning of the disease is currently of major consideration. Analysis of 459 soft-tissue sarcoma patients revealed that local recurrence rate decreased from 39% to 24% with an addition of post-operative radiotherapy. It was concluded that to improve the local control it is preferable to increase the use of radiotherapy with adequate margins instead of increasing surgical margin which cannot be achieved without increasing amputation rate (7).

The clinical target volume (CTV) boundary is the margin added to the radiographically visible gross tumor volume (GTV) which accounts for microscopic disease spread and defines the region receiving a high curative dose of radiation. To delineate the CTV, clinicians follow established guidelines for the lower extremities STS by creating a geometrical expansion of the GTV by 3 cm in the longitudinal (proximal and distal to the GTV) direction. Crosswise expansion of the GTV is smaller and does not exceed 1.5 cm (8, 9). The expansion is done by dilating the GTV contour and manually editing the CTV boundary on the CT scan used for treatment planning. Accuracy of the manual CTV delineation is limited by insufficient tissue contrast of the CT scan and limited visual perception of the 3D shape of the anatomy when using 2D views (10).

Microscopy studies show that sarcoma cells invade the muscle tissue by spreading preferentially along the muscle fibers (11, 12). Therefore, the boundary of the clinical target volume (CTV) has to be defined taking into account tissue anisotropy. Diffusion tensor MRI (DT-MRI) analysis allows to quantify the anisotropy of muscle structure and to determine muscle fiber orientation based on anisotropic diffusion of water molecules in the muscles (13).

It was established that the cancer is confined within the muscles it originates from. Indeed, myectomy, a surgical procedure of removing the entire involved muscle, has been proven to decrease local recurrence rate (14). Cancer spread is also confined by the natural anatomical barriers such as bones and fat. It is therefore desirable to segment anatomical images in order to identify tissue types and individual muscle boundaries for accurate definition of the CTV.

Previous DT-MRI-based methods have been used to define anisotropic CTV boundaries for glioma (15). Specifically, they used DTI tractography methods to predict trajectories of the tumor cell spread. In muscle tissue, DT-MRI has been extensively used in the context of orthopedic and sports medicine (16-18) but, to the best of our knowledge, not for modeling of tumor spread in soft tissue. In the present paper, we develop our model based on imaging of healthy volunteers to determine the feasibility of the promotion of DT-MRI for the target definition in clinical settings.

The innovation of our work lies in the derivation of muscle fiber orientation from DT-MRI *without resorting to tractography*, and in the application to anisotropic tumor spread modeling in STS. We expect that this work will become a useful first step towards automated delineation of the CTV in tissues with anisotropic properties.

**Materials and methods**

*Image acquisition*

Eight healthy volunteers, five men and three women, participated in this study which was approved by the Institutional Review Board of Massachusetts General Hospital. Written informed consent was obtained from each participant. The volunteers were scanned supine, feet first using 3T MRI system (Siemens, Magnetom Prisma, Siemens Healthcare, Erlangen, Germany) and an 18-channel phased array coil covering left and right thighs. The imaging protocol consisted of (a) two high resolution anatomical scans (spin-echo, SE), T1- and T2-weighted; (b) a diffusion weighted (DW) spin echo-based scan using an echo planar (EP) acquisition with fat suppression. Anatomical and diffusion-weighted MRI scans were acquired in the axial plane.

The DW-MRI acquisition consisted of two $b_0$ images with $b_0=50$ s/mm$^2$ and 12 DW images with b=400 s/mm$^2$ using 12 gradient directions. A spectral adiabatic inversion recovery (SPAIR) fat saturation was used to suppress the fat signal. 12 independent acquisitions were performed for each b=400 s/mm$^2$ diffusion gradient acquisition. The images without diffusion-weighting were independently acquired two times. T1- and T2-weighted acquisitions were used to match the anatomical location of the muscles in DW images. The other acquisition parameters are compiled in Table 1.

**Table 1.** Characteristics of MR Imaging, parameter (number of cases)

| Sequence | TR ms/TE ms | Spatial resolution, mm$^3$ | Number of slices | Reconstruction matrix | Acquisition time |
|---|---|---|---|---|---|
| DWI: EP | 7900/54 (3) 7900/78 (2) 10900/87 (1) 1600/46 (1) 3800/43 (1) | 1.25x1.25x6 (6) 3.125x3.125x5 (1) 1.5625x1.5625x1.6 (1) | 40 (6) 35 (1) 128 (1) | 980x2240 (6) 1320x3840 (1) 264x384 (1) | 22 min 33 s (5) 16 min 5 s (1) 30 min 42 s (1) 45 min 53 s (1) |
| T1-: SE | 218/17 (4) 9350/8.1 (2) 253/17 (1) 13670/9.2 (1) | 1x1x6.5 (7) 1x1x2 (1) | 32 (4) 30 (2) 40 (1) 100 (1) | 192x256 (5) 156x192 (3) | |
| T2-: SE | 250/8.5 (2) 9350/73 (2) 250/8.5 (2) 157/8.5 (1) 13670/74 (1) | 1x1x6.5 (7) 1x1x2 (1) | 32 (4) 30 (2) 40 (1) 100 (1) | 192x256 (5) 156x192 (3) | |

*Data processing and image segmentation*

The diffusion-weighted series were resampled to an isotropic voxel size of 1.25×1.25×1.25 mm$^3$. For each acquired DW-MRI scan, the diffusion tensor was reconstructed from 12 diffusion-encoded gradient pulses using imaging Python library DIPY (19) with the tensor model of Basser et al. (20). Based on the tensor image, scalar maps of the apparent diffusion coefficient (ADC) and fractional anisotropy (FA) were calculated.

Twelve muscles were manually contoured on T1-weighted MR scans of each left and right thigh (see the results in Fig. 1, Panel A). For automated segmentation of the muscles we trained the convolutional neural network (CNN) model with 2D U-Net architecture by Ronneberger et al. (21) using TensorFlow (22). Input of the network was T1-weighted MR volumes, apparent diffusion coefficient (ADC) volumes, and fractional anisotropy (FA) volumes (three channels) for 8 subjects (16 sets of 3 images of the thigh) which were cropped to the same size and resampled to isotropic resolution of 2×2×2 mm$^3$. For each channel, 100 2D slices from each image per thigh were used. The images were paired with the segmentation label to construct a multichannel sample (23, 24). The output was a mask with one channel for each muscle and one channel for unspecified tissue.

The network was trained with the pixel-wise cross entropy loss for 100 epochs, which was optimized with the Adam optimizer (alpha=0.001, beta1=0.9, beta2=0.999, and epsilon=$10^{-8}$). Drop-out layers at the end of the contracting path perform further implicit regularization as vanilla U-Net. The learning rate was 0.001. We used the typical image data augmentations of image crop, horizontal flip, Gaussian blur, and linear contrast change. The networks that used only MR image as input or only ADC or FA image as input differed from the above specification only by the number of input channels. We used leave-one-out validation with 5 subjects (1000 2D slices) for training and validated on 1 subject (200 2D slices) in each round, and rotated for 6 rounds. The image sets from 2 subjects (400 2D slices) were used for testing.

*Modeling CTV*

We assume that the cancer cells reach distant points following the shortest paths from the surface of the gross tumor volume (GTV) within the anisotropic surrounding tissues, and the CTV boundary can be thought of as a propagating front. The shape of the front depends on the spatial distribution of propagation speed, so at any given moment the front will end up at different distances from the GTV depending on the speeds. We will parametrize the fronts by the shortest distances, equal at every point on the front. Thus, we will consider the CTV boundary as isosurface of shortest distances.

In anisotropic tissue, such as muscle, we assume that preferential tumor spread occurs along muscle fibers. The directionality of the muscles is assessed by the water diffusivity in DW imaging which is the largest in the direction parallel to the dominant orientation of the fibers.

As a pre-processing step, we transform the diffusion tensors from the DT-MRI by keeping their eigenvectors but replacing the eigenvalues (in increasing order) by $\lambda_1 = \lambda_2 = 1$ and $\lambda_3 = 10$. One advantage of this approach compared with the direct use of the entire diffusion tensor is that by the choice of the $\lambda$ we can adapt the resulting CTV to clinical experience, and account for differences between water diffusivity and tumor cell spread.

To find the isosurfaces of shortest distances, we numerically solve the Eikonal equation in anisotropic media using an open source implementation of the Hamiltonian

Fast Marching Method (25) (26), see Appendix. To model the GTV we placed a sphere within the DT-MRI volume.

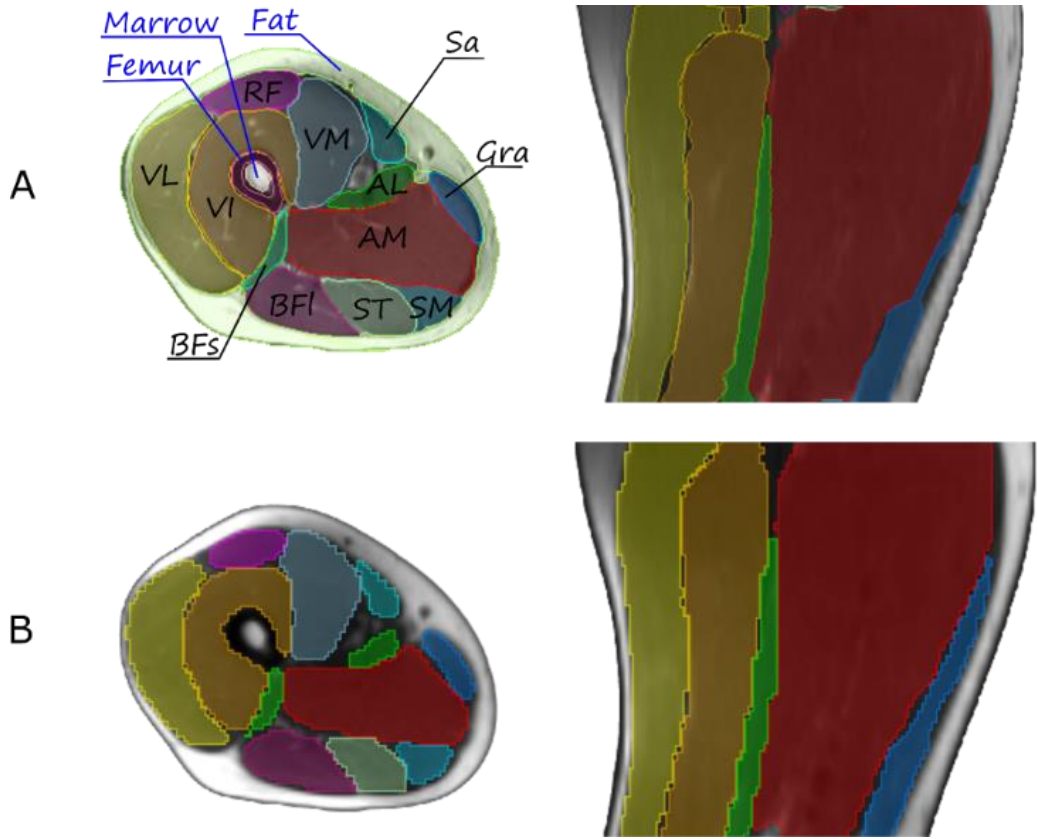

**Figure 1.** Representative T1-weighted MR anatomical image of the thigh, axial (left) and sagittal (right) views. The 12 muscles are: sartorius SAR, vastus medialis VM, vastus intermedius VI, vastus lateralis VL, rectus femoris RF, biceps femoris short head BFS, biceps femoris long head BFL, semitendinosus ST, gracilis GRA, semimembranosus SM, adductor longus AL, adductor magnus AM. Panel A: Manually segmented tissues and individual muscles. Panel B: results of automated segmentation of individual muscles.

## Results

*Imaging and image processing*

The image acquisition parameters and characteristics are listed in Table 1. A qualitative comparison of the manually delineated thigh muscles with the structures generated by the deep-learning model is shown in Fig. 1, Panels A, B. Quantitatively, the accuracy of the model was assessed with the Dice similarity coefficient (DSC) and presented with the boxplots in Fig. 2.

To investigate the effect of inclusion of the three imaging modalities (T1-weighted MRI, ADC, and FA maps) in the network training on the segmentation accuracy, we trained the model with each of these images separately as the input channels. The best accuracy was achieved with all three modalities included in the training. The most contributing input was from the T1-eighted MRI with the ADC and FA almost equally contributing, see the four panels in Fig. 2.

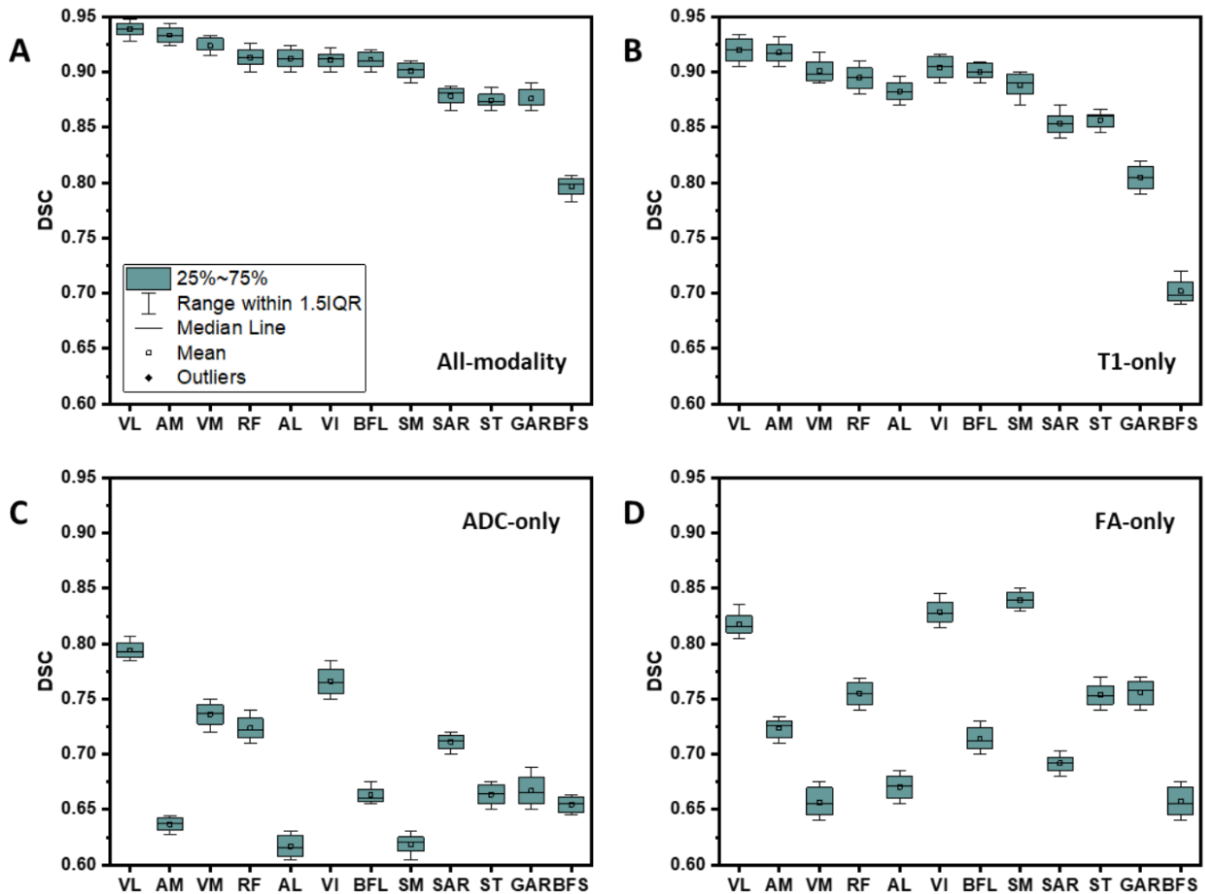

**Figure 2.** Boxplots of the Dice similarity coefficient (DSC) for automated segmentation as compared to the manual delineation. The inputs are A: three modalities, B: T1-weighted MR image only, C: ADC only, and D: FA only.

*DTI data-driven solution of the anisotropic Eikonal equation in muscles*

The voxel-wise diffusion tensors and the muscle masks were used with the Hamiltonian Fast-Marching solver to obtain the solution of Eq. (A.4) in the form of isosurfaces of the shortest distances when traveling from a point within the muscle.

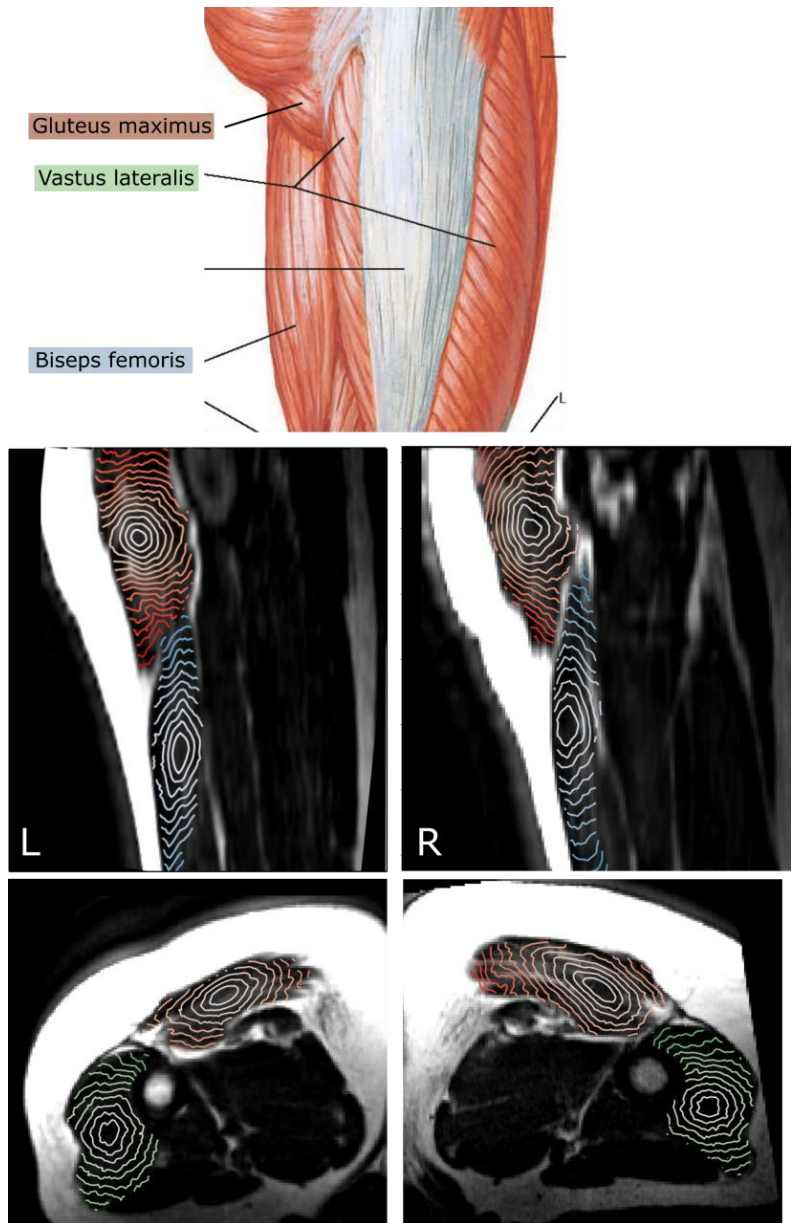

**Figure 3.** Top: muscles of the thigh, gluteus maximus, biseps femoris, vastus lateralis. (Adapted from (27)). Middle: sagittal view of the T1-weighted MRI with isosurfaces of shortest distance calculated within the gluteus maximus (orange) and biseps femoris (blue) muscles starting from a point in the center. Bottom: axial view with the isosurfaces calculated within gluteus maximus (orange) and vastus lateralis (green) muscles.

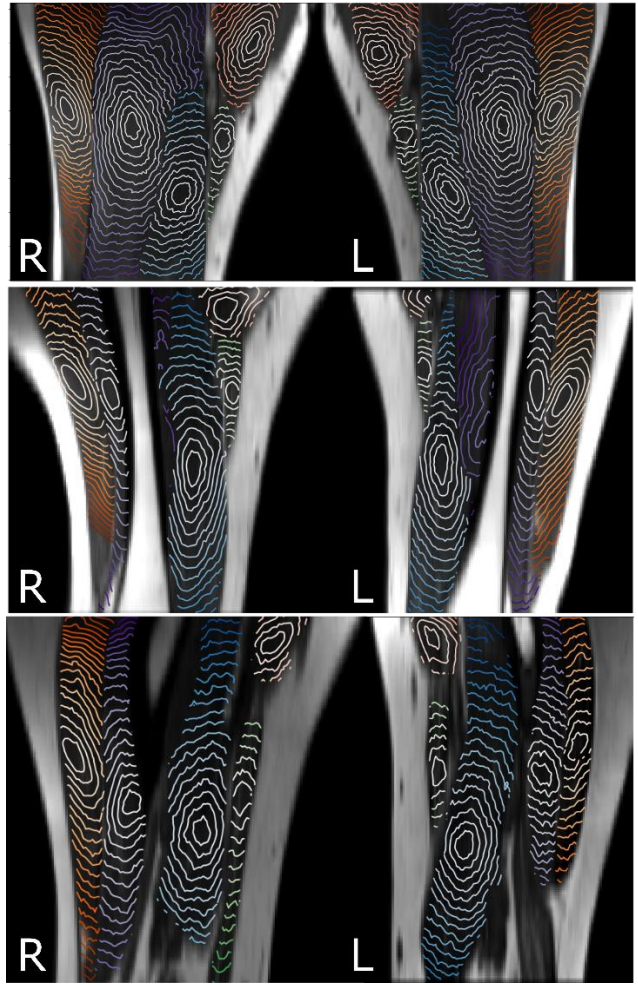

**Figure 4.** Isosurfaces of the shortest distance calculated with $\lambda_1 = \lambda_2 = 1$ and $\lambda_3 = 10$ within five muscles starting from a point in the center of vastus lateralis (orange), vastus intermedius (blue), vastus medialis (light blue), sartorius (green), and adductor longus (pink). Anisotropy of the isosurfaces is consistent in the left-right thigh and between three subjects.

In Fig. 3, we present the 2D sections of the isosurfaces calculated in the three muscles, gluteus maximus, biseps femoris, and vastus lateralis having anatomically different orientations with respect to each other in the thigh. Specifically, the biseps femoris and vastus lateralis are nearly parallel and both of them are nearly perpendicular to the gluteus maximus (see illustration in Fig. 3 adapted from (27)). The isosurfaces clearly show anisotropy of the muscle tissue which is consistent with the anatomical fiber orientation. Fig. 4 further demonstrates consistency of the DT-MRI in defining directionality of the muscle fibers. The examples show the isosurfaces calculated using data obtained from three subjects, both left and right thigh, with the isosurface spanning different muscles. The images show a high degree of consistency among all comparisons. For example, the anisotropy of the vastus lateralis muscle tissue (orange lines) is nearly identical in all six shown maps.

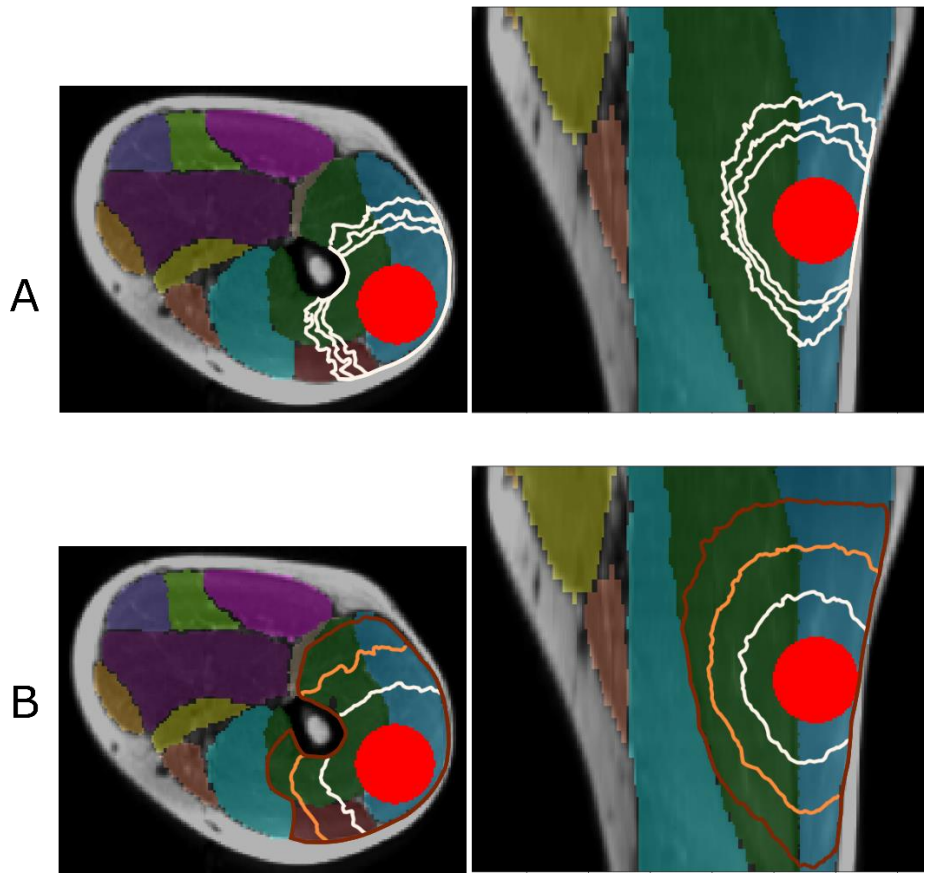

**Figure 5.** Panel A: isosurfaces of the shortest distance calculated in three muscles, vastus lateralis (blue shade), vastus intermedius (green), and rectus femoris (dark red) starting from the surface of modeled GTV with $\lambda_3 = 10$, $\lambda_3 = 20$, and $\lambda_3 = 40$. The outermost contour corresponds to the largest value of $\lambda_3$. Panel B: three levels of isosurfaces of shortest distance with $\lambda_3 = 10$. The boundary of the fat, femur, and non-involved muscles completes the modeled CTV.

To model the CTV in muscle tissue, we started by placing a model spherical GTV of 22.5 mm radius within vastus lateralis and vastus intermedius muscles. In Fig. 5 Panel A, the isosurfaces of shortest distance are shown for three values of the largest eigenvalue $\lambda_3$. It is expected that in the ideal case of a single dominant direction the asymmetry scales as $\sqrt{\lambda}$ (see phantom case experiments in Fig. A.1 in the Appendix). However, acquisition noise leads to variation of diffusion tensor eigenvalues and direction of its principal axes compared to their true values. As a result, the shape of isosurfaces becomes less elongated compared to the expected ratio of $\sqrt{\lambda}$.

For a given asymmetry, the distance from the GTV models a particular CTV boundary. Fig. 5 Panel B shows the candidate CTV boundaries obtained by solving Eq. (A.4) with averaged eigenvectors and $\lambda_3 = 10$, at the increasing distance cutoff. By varying the anisotropy $\lambda_3$ and distance cutoff, it is possible to tailor the shape of the automatically generated CTV to the satisfaction of the clinician.

**Discussion**

In this study, we leverage well established DT-MRI-based microstructural tissue characterization (13, 16, 17, 28, 29) combined with deep learning segmentation to automatically define potential CTV regions in the thigh muscles. Our method has an advantage over a previously developed DT-MRI-based method for defining anisotropic CTV boundary for glioma (15) which utilized DTI tractography to predict trajectories of the tumor cell spread. Since raw imaging data is used, the CTV definition no longer depends on some of the tractography modeling assumptions. Also, since we do not use the tensor eigenvalues and only the principal eigenvectors, the only underlying assumption of our CTV model is that the tumor cells preferentially spread along muscle fibers, in full agreement with recent microscopy experiments (11, 12).

Our approach has certain limitations. Because of the image directional noise, at least in the imaging protocol employed in this study, the isosurfaces are not smooth and eigenvalue $\lambda_3$ exerts a limited control over isosurface asymmetry. Approaches to reduce noise such as residual deep learning, low rank constraints, and wavelet could potentially further improve signal-to-noise ratio of diffusion-weighted images without incurring additional scan time and potentially allow for per pixel tensor calculation (30-32). Alternatively, this approach can be used to reduce scan time by reducing the number of signal averages as the DT-MRI scan time in this study is longer than would be acceptable for scanning cancer patient on a routine basis. Future work will include implementing such denoising techniques and evaluating its effect on the proposed CTV modeling. Lastly, since the study is performed on healthy volunteers, the proposed method needs to be validated using STS patient data. Specifically, the tissue at the site of the tumor can be altered leading to changes in diffusion measurements. Future studies will include analysis of clinical imaging data and comparison of the automatically generated CTV with the CTV contoured by radiation oncologists specializing in sarcoma.

**Conclusion**

We proposed and demonstrated preliminary feasibility of a novel approach of combining DT-MRI acquisition of the lower extremities with a CNN-based automatic segmentation to further refine proposed CTV boundaries on soft-tissue sarcoma. Future studies will be focused on the clinical validation and efficacy of the proposed technique in soft-tissue sarcoma patients.

**Appendix**

We model the boundary of the CTV as propagating front on the image voxel grid. In the isotropic case, the time $u$ to arrive from the surface of the GTV to a given point $x$ along a path $C(r)$, assuming scalar front propagation speed $v(r)$ is given by the line integral

$$u(x) = \int_C \frac{1}{v(r)} ds. \tag{A.1}$$

The earliest arrival time $u(x)$ along the shortest path $C_0(r)$ is obtained by solving the Eikonal equation,

$$\|\nabla u(x)\| = \frac{1}{v(x)}, \qquad u(x)|_{\partial \Omega} = 0, \tag{A.2}$$

where $u(x)$ is the front arrival time when traveling from the boundary $\partial \Omega$ (the surface of the GTV) to the point $x$ with the speed $v(x)$, and $\|\cdot\|$ is the Euclidean norm. In the context of the CTV, it is more natural to think of shortest distances $S(x)$ rather than earliest arrival times $u(x)$. By multiplying both sides of Eq. (A.2) with $v_0$ defined as a reference speed of tumor cell propagation in soft tissue we arrive at the equation

$$\|\nabla S(x)\| = \frac{v_0}{v(x)}, \qquad S(x)|_{\partial \Omega} = 0. \tag{A.3}$$

The parameter $v_0$ can be adjusted to fit tumor progression data as it becomes available.

In an intrinsically anisotropic case the front propagation speed is not only spatially variable, but also depends on the direction of propagation. Front propagation is then described by a positive definite matrix $\widehat{M}$, or Riemannian metric (33, 34). In the basis that diagonalizes the matrix $\widehat{M}$, its diagonal elements (eigenvalues $\lambda_1, \lambda_2, \lambda_3$) are inversely proportional to the squares of the propagation speed along the principal directions. As a consequence, the iso-distant surfaces of $S(x)$ scale with $1/\sqrt{\lambda}$. In this anisotropic case the shortest distance $S(x)$ is given by the solution of the anisotropic Eikonal equation,

$$\|\nabla S(x)\|^2_{\widehat{M}^{-1}(x)} = \nabla^T S(x) \cdot \widehat{M}^{-1}(x) \cdot \nabla S(x) = 1, \qquad S(x)|_{\partial\Omega} = 0. \tag{A.4}$$

In our model, we expect that larger water diffusivity corresponds to earlier arrival times and smaller values of $S(x)$. Thus, we use the tensor of water diffusivity $\widehat{D}$ as $\widehat{M}^{-1}$ when solving (A.4). Here the iso-distant surfaces of $S(x)$ scale with the squareroot of the eigenvalues of $\widehat{D}$.

We tested the method using a phantom case simulating a voxel volume of size 256×256×256 by reproducing simple geometrical shapes. The voxel values are 3×3 diagonal tensors with eigenvalues $\lambda_1$, $\lambda_2$ and $\lambda_3$ and eigenvectors $\vec{V}_1$, $\vec{V}_2$, and $\vec{V}_3$ aligned with the voxel volume's axes X, Y, and Z as illustrated in Fig. A.1. The relative values of $\lambda_1$, $\lambda_2$ and $\lambda_3$ determine directional properties of the media within the volume. In isotropic media, $\lambda_1 = \lambda_2 = \lambda_3 = 1$ and the calculated map of the shortest path lengths is a series of spherical surfaces (Panel A in Fig. A.1). We introduce anisotropy by weighting voxel-wise Z-direction such as $\lambda_1 = \lambda_2 = 1$ and $\lambda_3 = 5$. In this case, the calculated iso-distance surfaces are ellipsoids (Panel B in Fig. A.1). The degree of anisotropy defined as a ratio of the principal axes of the ellipsoid is $a = \sqrt{5} \approx 2.24$. With increasing anisotropy, for $\lambda_3 = 50$, the degree of anisotropy increases, $a = \sqrt{50} \approx 7.1$, approaching a thin "cigar" shape (Panel C in Fig. A.1).

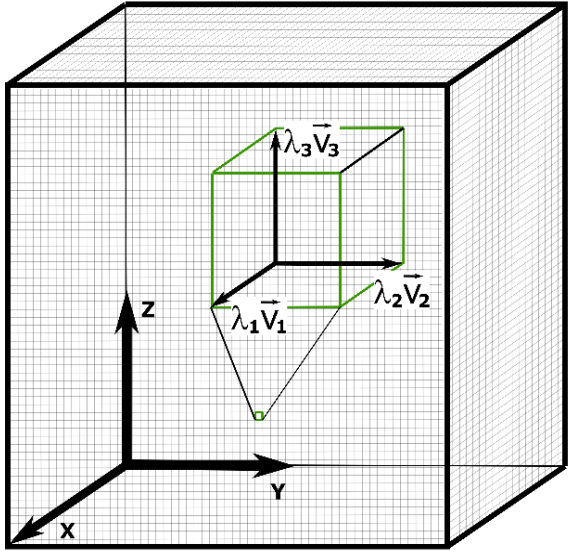

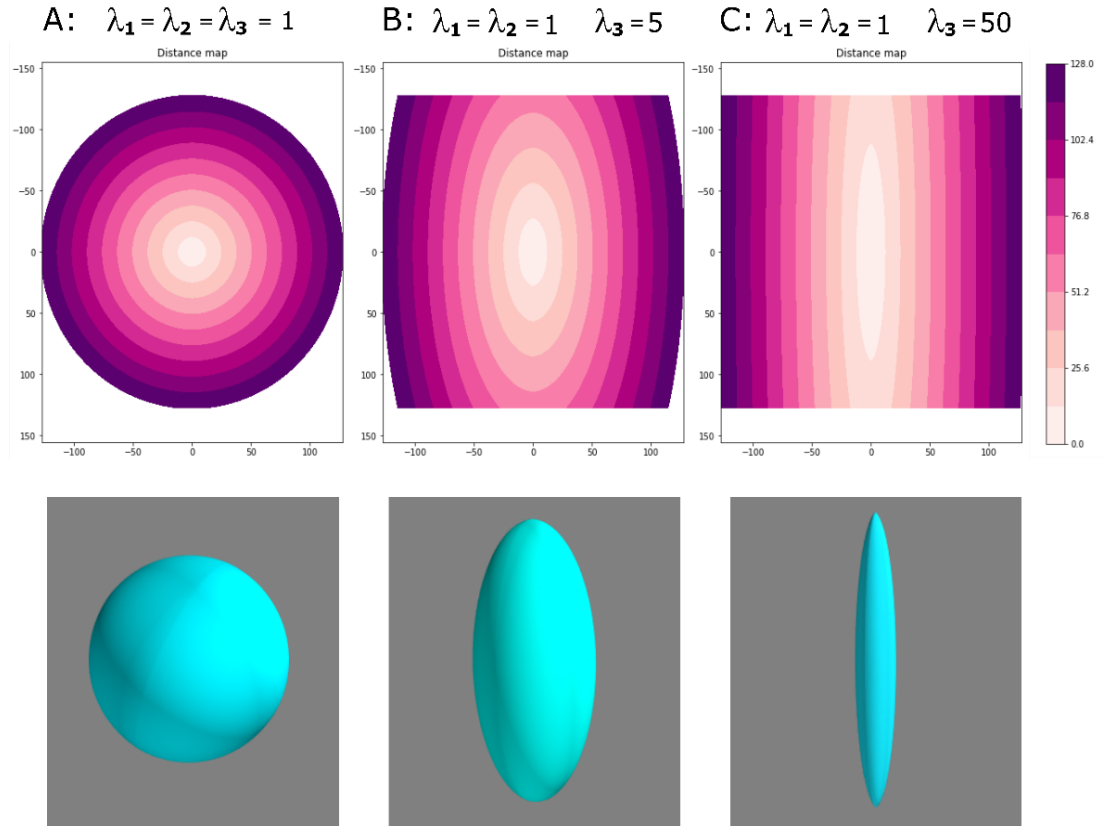

**Figure A.1.** Schematic representation of the image volume with the coordinate axes, X (anterior-posterior), Y (left-right), Z (inferior-superior) and voxel-wise diagonal tensor with eigenvalues $\lambda_1$, $\lambda_2$ and $\lambda_3$ and eigenvectors $\vec{V}_1$, $\vec{V}_2$, and $\vec{V}_3$ along the image axes. Lateral cross-sections of the iso-surfaces in the X-Z plane, and 3D renderings of a representative iso-surface are shown for $\lambda_1 = \lambda_2 = 1$ and $\lambda_3 = 1,5,50$ (left to right).